\documentclass[preprint]{JHEP} % 10pt is ignored!
\usepackage{epsfig}

\newcommand{\nco}{\newcommand}
\nco{\beq}{\begin{equation}}
\nco{\eeq}{\end{equation}}
\nco{\beqa}{\begin{eqnarray}}
\nco{\eeqa}{\end{eqnarray}}
\nco{\lra}{\leftrightarrow}
\def\sfrac#1#2{{\textstyle{#1\over #2}}}
\def\eps{\epsilon}
\nco{\sss}{\scriptscriptstyle}
\nco{\lsim}{\mbox{\raisebox{-.6ex}{~$\stackrel{<}{\sim}$~}}}
\nco{\gsim}{\mbox{\raisebox{-.6ex}{~$\stackrel{>}{\sim}$~}}}

\title{Order $\mathbf\rho^2$ Corrections to Randall-Sundrum ${\mathbf I}$ 
 Cosmology}
\author{by James M.\ Cline and J\'er\'emie Vinet\\
Physics Department, McGill University, Montr\'eal, Qu\'ebec, Canada H3A
2T8\\
E-mail: \email{jcline@physics.mcgill.ca},
\email{vinetj@physics.mcgill.ca}}
\preprint{McGill 02-01}
\abstract{We derive the corrections to the Friedmann equation of order $\rho^2$
in the Randall-Sundrum (RS) model,  where two 3-branes bound a slice of
five-dimensional Anti-deSitter space.  The effects of radion stabilization by
the Goldberger-Wise mechanism are taken into account. 
Surprisingly, we find that an inflaton on either brane will experience no
order $\rho^2$ corrections in the Hubble rate $H$ due to its own energy
density, although an observer on the opposite brane does see such a 
correction.  Thus there is no enhancement of the slow-roll condition unless
inflation is simultaneously driven by inflatons on both branes.
Similarly, during radiation domination, the $\rho^2$ correction to $H$ on
a given brane vanish unless there is nonvanishing energy density on the opposite
brane.  During the electroweak phase transition the correction can be large,
but is has the wrong sign for causing sphalerons to go out of thermal
equilibrium, so it cannot help electroweak baryogenesis.  We discuss the
differences between our results and exact solutions in RS-II cosmology.}

\begin{document}

\section{Introduction}

The idea of our world being a 3-brane embedded in a higher dimensional
bulk has provided an attractive explanation for the weakness of gravity
compared to the other fundamental forces \cite{ADD,RSI,RSII}.  In this
framework, 4-D gravity is weak at distances much greater than the
fundamental Planck length, usually assumed to be the TeV$^{-1}$ scale, 
but it quickly becomes strong at shorter distances.  This idea has rich
consequences for LHC physics, and it also implies modifications to
cosmology at temperatures approaching 1 TeV.  Although it may not be
feasible to test deviations from standard cosmology which happened so
early, this scale is sufficiently close to the electroweak phase
transition to give some hope of observable consequences, for example to
electroweak baryogenesis.  Moreover the nature of inflation can be
significantly altered by extra-dimensional effects \cite{inf,MWBH,CLL,APR}.

A seminal work on brane cosmology \cite{BDL} observed that the Friedmann
equation on a 3-brane embedded in a flat 5-D spacetime takes the form $H^2
= {\rho^2\over 36 M_5^6}$ instead of the standard form $H^2 = {8\pi G\rho\over
3}$.  
Here $M_5$ is the 5-D Planck mass and $\rho$ is the 4-D energy density on
the brane.  This strange result arises from the fact that the brane is
assumed to be infinitesimally thin compared to the size of the extra
dimension ($y$), which introduces a discontinuity in the derivatives of
metric elements with respect to $y$ at the position of the brane.  The
$\rho^2$ dependence is strongly contradicted by big bang nucleosythesis.  
It was subsequently observed \cite{CGKT,CGS} that by adding a negative
cosmological constant $\Lambda_5$ to the 5-D bulk, and a tension ${\cal T}$
to the brane, the Friedmann equation generalizes to $H^2 =
{\Lambda_5\over 6M_5^3}  + {({\cal T}+\rho)^2\over 36 M_5^6}$, 
which takes the form
$H^2 = {8\pi G\rho\over 3}(1+ O({\rho\over M_5^4}))$ if one chooses ${\cal T}^2 =
-6\Lambda_5 M_5^3$ and ${{\cal T}\over 18 M_5^6} = {8\pi G\over 3}$.  In other words,
we recover the standard Friedmann equation, modified by $\rho^2$
corrections which are small as long as $\rho\ll {\cal T}\sim$ TeV$^4$.  In
addition, the two relations between ${\cal T}$ and $M_5$ are equivalent to
those needed for the Randall-Sundrum solutions, where the bulk is a region
of 5-D Anti-DeSitter space.

Although the problem of how to recover the correct Friedmann equation was
partially resolved, refs.\ \cite{CGKT,CGS} pointed out that in the
two-brane model of \cite{RSI} (RS-I), $H^2$ could be real only if $\rho$
was assumed to be negative on the second brane, whose tension is negative.
In addition one had to assume a fine-tuned relation between the energy
densities on the two branes.  This untenable result was especially
troubling because the negative tension, or TeV brane, is the one where we
are assumed to be living in order to perceive 4-D gravity as being weak. 
Ref.\ \cite{KKOP} pointed out that these problems could be avoided if
the $T_{55}$ component of the bulk stress-energy tensor had the right
properties. 
It was subsequently shown \cite{CGRT,KKOP2} that stabilization of the size of
the extra dimension (the radion mode) was necessary and sufficient for
recovering acceptable cosmology on the TeV brane.  The unphysical
requirement of a fine-tuned negative value for $\rho$ was merely a
consequence of assuming the extra dimension was static when generically,
in the absence of any stabilizing force, it would tend to expand or
contract.  The positive effects of stabilization were subsequently
confirmed in greater detail by references \cite{MT,CGK} and \cite{CF}.
The problem of losing stabilization at temperatures above the TeV scale 
and tunneling back to the stabilized ground state was explored in
ref.\ \cite{CF1}.

The issue of radion stabilization exists only in the RS-I and ADD
\cite{ADD} solutions; in RS-II \cite{RSII} the extra dimension is infinite
in size, there is only a single brane, and there is no radion---its wave function
diverges away from the brane, and so it is not a normalizable mode if there is
no second brane to compactify the space.  Because of
this simplicity, much of the early work on brane cosmology has focused on
the RS-II solution.  In this case the above-quoted form of the modified
Friedmann equation becomes {exact},
\beq
\label{mfe}
	H^2 = {8\pi G\over3}\rho\left(1+ {\rho\over 2{\cal T}}\right)
\eeq
as long as higher-derivative corrections to the gravitational action are
neglected\footnote{Since the discontinuity in the derivative of the warp
factor at the brane is proportional to $\rho$ as a consequence of the 5-D
Einstein equations, one may expect that higher derivative terms will bring
higher powers of $\rho$.} and the bulk is empty.  Perhaps the most important application of the
$\rho/{\cal T}$ correction is to increase the expansion rate over its
normal value during inflation \cite{MWBH,CLL}.  This allows for the possibility
of fulfilling the slow-roll condition with steeper inflationary potentials
than would normally be admissible, with the possible observable consequence
of a spectral index of density perturbations which is smaller than in
conventional models of inflation.  
It should be kept in mind that the effects of the $\rho/{\cal T}$
correction should not be trusted quantitatively whenever they become large,
precisely because of the neglect of higher derivative operators which are
only suppressed by the energy scale $\sim {\cal T}^{1/4}$.  Beyond this
scale one really needs the full quantum theory of gravity, presumably
string theory, to make quantitative predictions.  Thus the $\rho/{\cal T}$
corrections should be regarded as indicative of the kinds of new
qualitative effects that could be expected at energy densities above the
quantum gravity scale.

More recently, the cosmology of the RS-I model has attracted renewed
attention.   The AdS/CFT correspondence has been used as a way of better
quantifying the quantum gravity effects: Kaluza-Klein excitations of the
bulk graviton modes are supposed to be equivalent to bound states of a
strongly coupled, nearly conformal field theory residing on the TeV brane
\cite{Gubser}.   At temperatures above the TeV scale, the TeV brane
is supposed to be hidden behind a horizon \cite{APR} associated with a
black hole which formed in the bulk \cite{HMR}.  It has been noted that
the emergence of the TeV brane from the horizon may occur around the same
time as the electroweak phase transition \cite{CNR}.  Ref.\ \cite{CF1}
provided an alternative picture of a first order phase transition which
leads to the appearance of the TeV brane at this epoch.

There have been numerous investigations of the effects of a bulk scalar
field on brane-world cosmology \cite{scalar}.
In the present work we shall be concerned with the $O(\rho^2)$ corrections
to the Friedmann equation which arise from the 5-D nature of the underlying
geometry in the RS-I model, using the Goldberger-Wise (GW) \cite{GW2} mechanism
to stabilize the size of the extra dimension.  In general one no longer
expects eq.\ (\ref{mfe}) to hold in the RS-I model because of the
additional effect of the dynamical radion.  The radius is displaced by the
cosmological expansion, which changes the strength of gravity and thus
back-reacts on the expansion rate.  Although one should not trust these
corrections quantitatively when they start to become large (for the reasons
discussed above), they accurately predict the deviations from standard
cosmology as one starts to approach the quantum gravity scale around 1 TeV.

The plan of the paper is as follows.  In section 2 we expand the 5-D Einstein
equations to second order in a perturbation series in $\rho$ and $\rho_*$, the
energy densities on the TeV and Planck branes.  In section 3 the equations are
solved to find the ingredients from which one can infer the effective 4-D
Friedmann equations.   In section 4 we present the results for the Hubble rate and
the acceleration as measured on either brane, for arbitrary equations of state,
and we discuss the implications for cosmology.  We also compare the results to
where there is only a single brane or where there is no bulk scalar. The reader
who is not interested in the details of solving the Einstein equations can go
directly to this section.  Section 5 gives a summary and conclusions.  Some
technical details can be found in the appendices.

\section{$\mathbf O(\rho^2)$ Einstein Equations }

Following closely the formalism of ref.\ \cite{CF}, we will look for
cosmological solutions to 5-D gravity coupled to the Goldberger-Wise 
scalar field $\Phi$, which is responsible for stabilizing the radion, and
to two branes: the Planck brane, located at $y_0\equiv 0$, and the TeV
brane at $y_1\equiv 1$. The action is
\beqa
S &=& \int d^{\,5}x\sqrt{g} \left( -{1\over 2\kappa^2}R - \Lambda + 
    \sfrac{1}{2}\partial_{m}\Phi\partial^{m}\Phi
        -V(\Phi)\right)\nonumber\\
        &+&\int d^{\,4}x \sqrt{g}\left({\cal L}_{m,0} -
V_{0}(\Phi)\right)|_{y=0}
        +\int d^{\,4}x \sqrt{g}\left({\cal L}_{m,1} -
        V_{1}(\Phi)\right)|_{y=1},
\eeqa
where $\kappa^2=M_5^{-3}$, and the potential $V(\Phi)$ is left unspecified for now.
%\beq
%	V(\Phi) = \sfrac12 m^2\Phi^2;\qquad
%	V_i(\Phi) = {\cal T}_i + \lambda_i(\Phi^2 - v_i^2)^2.
%\eeq
The brane contributions are a sum of matter, represented by 
\beq
	{\cal L}_{m,0}\sim\rho_*;\qquad  {\cal L}_{m,1}\sim\rho
\eeq
and tension, which is the value of the brane's scalar
potential $V_i(\Phi(y_i))$.  The matter Lagrangians cannot be written
explicitly for cosmological fluids, but their effect on the Einstein
equations  is specified through
their stress-energy tensors (\ref{stress}).
Our ansatz for the 5-D metric has the form
\beqa
\label{ansatz}
	ds^{2} &=& n^2(t,y)\,dt^{2} 
        -a^2(t,y)\sum_i dx_i^2 -b^{2}(t,y)\,dy^{2}
	\nonumber \\
               &\equiv& e^{-2N(t,y)}dt^2-
		a_0(t)^{2}e^{-2A(t,y)}
		\sum_i dx_i^2 -b^2(t,y)\,dy^{2}.
\eeqa
We will make a perturbative expansion in the energy densities
$\rho,\rho_*$ of
the branes around the static solution, where $\rho=\rho_*=0$:
\beqa
		N(t,y) &=& A_0(y) + \delta N_1(t,y) + \delta N_2(t,y); \qquad
		\!\!\!
		A(t,y) = A_0(y) + \delta A_1(t,y) + \delta A_2(t,y) \nonumber\\
		b(t,y) &=& b_0 + \delta b_1(t,y)+ \delta b_2(t,y) ; \qquad\qquad
		\Phi(t,y) = \Phi_0(y) + \delta\Phi_1(t,y) + \delta\Phi_2(t,y).
		\nonumber\\
\eeqa
The subscripts on the perturbations indicate their order in powers of 
$\rho$ or $\rho_{*}$, both of which are taken to be formally of the same order for the
purposes of developing the perturbation expansion, although in actual order
of magnitude we may consider them to have very different values.
This ansatz is to be substituted into the scalar field equation,
\begin{equation}
	\partial _{t}\left(\frac{1}{n}ba^{3}\dot{\Phi}\right)
	-\partial_{y}\left(\frac{1}{b}a^{3}n\Phi^{\prime}\right)
	+a^{3}n\left[ b V'
	+ V_{0}'\delta(y)
	+ V_{1}'\delta(y-1) \right] =0,
%	+ba^{3}n\left[ \frac{\partial V}{\partial \Phi}
%	+\frac{\partial V_{0}}{\partial	\Phi}\delta(by)
%	+ \frac{\partial V_{1}}{\partial \Phi}\delta(b(y-1)) \right] =0,
\end{equation}
and into the Einstein equations, $G_{mn}= \kappa^2 T_{mn}$.
Here and in the following, primes on functions of $y$ denote
${\partial\over \partial y}$, while primes on potentials of $\Phi$
will mean ${\partial\over \partial \Phi}$.
The components of the Einstein tensor which do not vanish identically are
\beqa
\label{EE}
G_{00} &=& 3\left[\left(\frac{\dot {a}}{a}\right)^{2}+\frac{\dot {a}}{a}
	\frac{\dot {b}}{b}-\frac{n^{2}}{b^{2}}\left(\frac {a^{\prime \prime}}{a}
	+\left(\frac{a^{\prime}}{a}\right)^{2}
	-\frac{a^{\prime}b^{\prime}}{ab}\right)\right]\nonumber\\
G_{ii}&=&\frac{a^{2}}{b^{2}}
	\left[\left(\frac{a^{\prime}}{a}\right)^{2}
	+2\frac{a^{\prime}}{a}\frac{n^{\prime}}{n}
	-\frac{b^{\prime}}{b}\frac{n^{\prime}}{n}-	
	2\frac{b^{\prime}}{b}
	\frac{a^{\prime}}{a}+2\frac{a^{\prime
	\prime}}{a}+\frac{n^{\prime \prime}}{n}\right]\nonumber\\
	&+&\frac{a^{2}}{n^{2}}\left[-\left(\frac{\dot
	{a}}{a}\right)^{2}+2\frac{\dot{a}}{a}\frac{\dot{n}}{n}
	-2\frac{\ddot{a}}{a}+\frac{\dot{b}}{b}\left(-2\frac{\dot{a}}{a}
	+\frac{\dot{n}}{n}\right)-\frac{\ddot {b}}{b}\right]\nonumber\\
G_{05}&=&3\left[\frac{n^{\prime}}{n}\frac{\dot{a}}{a}
	+\frac{a^{\prime}}{a}\frac{\dot{b}}{b}
	-\frac{\dot{a}^{\prime}}{a}\right]\nonumber\\
G_{55}&=&3\left[\frac{a^{\prime}}{a}\left(\frac{a^{\prime}}{a}
	+\frac{n^{\prime}}{n}\right)-\frac{b^{2}}{n^{2}}
	\left(\frac{\dot{a}}{a}\left(\frac{\dot{a}}{a}
	-\frac{\dot{n}}{n}\right)+\frac{\ddot{a}}{a}\right)\right]
\eeqa
and the stress energy tensor is
$T_{mn} = g_{mn}(V(\Phi)+\Lambda)+\partial_{m}\Phi\partial_{n}\Phi
-\frac{1}{2}\partial^{l}\Phi\partial_{l}\Phi g_{mn}$ in the bulk.
On the branes, $T_m^{\ n}$ is given by
\begin{eqnarray}
T_{m}^{\ n} &=&{\delta(y)\over b(t,0)} \,{\rm 
diag}(V_{0}+\rho_{*},V_{0}-p_{*},V_{0}-p_{*},V_{0}-p_{*},0)
\nonumber \\  \label{stress}
&+&{\delta(y-1)\over b(t,1)}\,{\rm diag}(V_{1}+\rho,V_{1}-p,V_{1}-p,V_{1}-p,0)
\end{eqnarray}
(Later we will assume that the potentials $V_0$ and $V_1$ are very stiff
and are vanishing at their minima, so they can be neglected.)

The perturbative solution has been given to first order in $\rho,\rho_*$
in \cite{CF}; here we want to extend the analysis to second order.
Following \cite{CF}, it is useful to work with the following linear
combinations of the metric tensor perturbations:
\beq
	\Psi_2 = \delta A'_2 - A'_0 {\delta b_2\over b_0}-\frac{\kappa^2}{3} \Phi_0' \delta \Phi_2-\frac{\kappa^2}{6}
	\left(\delta \Phi_1'+\Phi_0' 
	{\delta b_1\over b_0}\right) \delta \Phi_1;\qquad
	\Upsilon_2 = \delta N'_2 - \delta A'_2.
\eeq
These are convenient variables because they appear in a natural way in the
boundary conditions at the two branes.  By integrating the (00) and (ii)
Einstein equations in the vicinity of either brane, we obtain the perturbed
Israel junction conditions,
\beqa
\label{Psibc}
\Psi_2(t,0) &=&
 +\left.\frac{\kappa^{2}}{6}b_{0}\rho_{*} {\delta b_1\over
b_0}\right|_{t;\,y=0} ;\qquad\qquad
\Psi_2(t,1)=-\left.\frac{\kappa^{2}}{6}b_{0}\rho{\delta b_1\over b_0}
\right|_{t;\,y=1}\\
\label{Upsbc}
 \Upsilon_2(t,0) &=&
 -\left.\frac{\kappa^{2}}{2}b_{0}(\rho_{*}+p_*){\delta
b_1\over b_0}\right|_{t;\,y=0};\quad 
\Upsilon_2(t,1)= 
+\left.\frac{\kappa^{2}}{2}b_{0}(\rho+p){\delta b_1\over
b_0}\right|_{t;\,y=1}
\eeqa
The analogous quantities at first order were found to be
\beq
\label{PsiUpseqs}
	\Psi_1 = \delta A' - A'_0 {\delta b_1\over b_0}-\frac{\kappa^2}{3} 
	\Phi_0' \delta\Phi_1;\qquad
	\Upsilon_1 = \delta N'_1 - \delta A'_1
\eeq
in \cite{CF}, and they satisfy the same boundary conditions as in
(\ref{Psibc},\ref{Upsbc}), but with the replacement ${\delta b_1\over b_0}\to 1$.

The Einstein equations take a form resembling ordinary second order
differential equations in $y$ for the variables $\Psi_2$
and $\Upsilon_2$, where all the time dependence of the latter is implicit,
through $\rho(t)$, $\rho_*(t)$, $p(t)$ and $p_*(t)$.  The quantities ${\dot
a_0\over a_0}$ and
${\ddot a_0\over a_0}$ function like constants of integration in this context,
whose values are fixed by imposing the boundary conditions
(\ref{Psibc},\ref{Upsbc}).  This procedure results in o.d.e.'s in time for
$a_0(t)$, which are the Friedmann equations we are seeking.  Accordingly,
we need to expand  ${\dot a\over a}$ and ${\ddot a\over a}$ in powers of
$\rho$.  Recall that in ordinary cosmology, $({\dot a\over a})^2 \sim
{\ddot a\over a} \sim \rho$.  Therefore each time derivative of $a_0$
counts as half a power of $\rho$.  In the brane cosmology we have $({\dot
a\over a})^2 \sim {\ddot a\over a} \sim \rho,\rho_* + O(\rho_*^2,\rho\rho_*,\rho^2)$, 
so we expect that
\beqa 
\label{1storderH}
{\dot a_0\over a_0} &=&  \sqrt{{8\pi G\over 3}\left(\rho_*+\Omega^4\rho\right)}
\ \Bigl[1 + O(\rho_*,\rho) \Bigr]
	\equiv \left({\dot a_0\over
a_0}\right)_{\!\!\!\left(\sfrac12\right)}
	+ \left({\dot a_0\over a_0}\right)_{\!\!\!\left(\sfrac32\right)} +
		\dots\nonumber\\
\left({\dot a_0\over a_0}\right)^2 &=& {8\pi G\over
3}\left(\rho_*+\Omega^4\rho\right)\Bigl[1 + O(\rho_*,\rho) \Bigr]          
        \equiv \left({\dot a_0\over a_0}\right)^2_{\!\!(1)} +
         \left({\dot a_0\over a_0}\right)^2_{\!\!(2)} + \dots
\eeqa 
where we have used subscripts to denote the power of $\rho$ or $\rho_*$
contained in the associated term.  Here $\Omega\equiv e^{-A_0(1)}$ is the
value of the warp factor at the TeV brane, which solves the hierarchy
problem in the RS scenario, and the linear combination $\rho_*+\Omega^4\rho$
is the result that has been proven \cite{CGRT, CF} to appear in the lowest order Friedmann equation
when the radion is stabilized.
We therefore expand ${\dot a\over a}$ and ${\ddot a\over a}$ in the
following manner:
\beqa
\label{Feq1}
\left({\dot a\over a}\right)^2 &=& \left({\dot a_0\over a_0} - \dot A\right)^2
\quad = \quad\left({\dot a\over a}\right)^2_{\!\!(1)} 
+ \left({\dot a\over a}\right)^2_{\!\!(2)} + O (\rho^3)\nonumber\\
&=& \left(\left({\dot a_0\over
a_0}\right)_{\!\!\!\left(\sfrac12\right)}\right)^2 + \left[2\left({\dot
a_0\over
a_0}\right)_{\!\!\!\left(\sfrac12\right)}\left({\dot a_0\over
a_0}\right)_{\!\!\!\left(\sfrac32\right)} - 2\left({\dot a_0\over
a_0}\right)_{\!\!\!\left(\sfrac12\right)}
\delta {\dot A_1}\right]+ O
(\rho^3);\nonumber\\
&=& \left({\dot a_0\over a_0}\right)^2_{\!\!(1)}  + \left[\left({\dot a_0\over a_0}\right)^2_{\!\!(2)} - 2\left({\dot a_0\over
a_0}\right)_{\!\!\!\left(\sfrac12\right)}
\delta {\dot A_1}\right]+ O
(\rho^3);\\
\label{Feq2}
{\ddot a\over a} &=&
{\ddot a_0\over a_0}- 2{\dot a_0\over a_0}{\dot A}-\ddot A + \dot A^2
\quad =\quad \left({\ddot a\over a}\right)_{\!\!(1)} + \left({\ddot a\over
a}\right)_{\!\!(2)}+ O (\rho^3)\nonumber\\
&=& \left({\ddot a_0\over a_0}\right)_{\!\!(1)} + \left[\left({\ddot
a_0\over
a_0}\right)_{\!\!(2)} -
2\left({\dot a_0\over a_0}\right)_{\!\!\!\left(\sfrac12\right)} \delta
{\dot
A_1} -\delta {\ddot A_1}\right]+ O (\rho^3).
\eeqa

In terms of these variables, we can write the second order Einstein 
equations (specifically, the linear combinations (00), $(00)+\frac{n^2}{a^2}(ii)$, (05) and (55), in that
order) as
\beqa
\label{00eq} %%%%%%%%%%%%%%%%%%%%%%%%%%%%%%%%%%%%%%%%%%%%%%%%%%%%%%%%%%%
	 \left({\dot a_0\over a_0}\right)_{\!\!(2)}^2 b_0^2 e^{2A_0} &=&
4 A_0' \Psi_2 - \Psi'_2 +{\cal F}_{\Psi} \\
\label{00iieq} %%%%%%%%%%%%%%%%%%%%%%%%%%%%%%%%%%%%%%%%%%%%%%%%%%%%%%%%%
	2\left( \left({\dot a_0\over a_0}\right)_{\!\!(2)}^2 -
 	\left({\ddot a_0\over a_0}\right)_{\!\!(2)}\right) 	b_0^2 e^{2A_0} 
	&=&- 4 A_0'\Upsilon_2  + \Upsilon'_2+{\cal F}_{\Upsilon} \\
\label{05eq} %%%%%%%%%%%%%%%%%%%%%%%%%%%%%%%%%%%%%%%%%%%%%%%%%%%%%%%%%
	0 &=& - \left({\dot a_0\over
	a_0}\right)_{\!\!\!\left(\sfrac12\right)}\!\!\Upsilon_2 + \dot\Psi_2
	+{\cal F}_{05}\\
\label{55eq} %%%%%%%%%%%%%%%%%%%%%%%%%%%%%%%%%%%%%%%%%%%%%%%%%%%%%%%%%%%
\left( \left({\dot a_0\over a_0}\right)_{\!\!(2)}^2 +
       \left({\ddot a_0\over a_0}\right)_{\!\!(2)}\right)b_0^2 e^{2A_0} &=&
	A_0'(4\Psi_2 + \Upsilon_2) + \frac{\kappa^2}{3}\left( \Phi_0''\delta
	\Phi_2-\Phi_0'\delta
	\Phi_2'+\Phi_0'^{2}\frac{\delta b_2}{b_{0}}\right) + {\cal F}_{55}\nonumber\\
	%&&\qquad\qquad\qquad\qquad\qquad\qquad+ {\cal F}_{55}	
\eeqa
and the scalar field equation as
\beqa
\label{scalar}%%%%%%%%%%%%%%%%%%%%%%%%%%%%%%%%%%%%%%%%%%%%%%%%%%%%%%%%%%%
\delta\Phi_2'' =
&&(4\Psi_2+\Upsilon_2)\Phi_0'+\left(\frac{4\kappa^2}{3}\Phi_0'^2+b_0^2
V''(\Phi_0)\right)\delta\Phi_2 +\left(2b_0^2
V'(\Phi_0)+4A_0'\Phi_0'\right)\frac{\delta b_2}{b_{0}}\nonumber\\ &&+\Phi_0'
\frac{\delta b_2}{b_{0}}'+{\cal F}_{\Phi}
\eeqa
where all the dependence on first order quantities squared is contained in the
functions ${\cal F}_{\Psi}$, ${\cal F}_{\Upsilon}$, ${\cal F}_{05}$, 
${\cal F}_{55}$ and ${\cal F}_{\Phi}$, which can be found in appendix A.

We have used the symbolic manipulation features of Matlab and Maple to arrive at this
and the other results in this paper, giving us confidence that there are no
algebraic errors, despite the complexity of the formulas. We emphasize
that all the terms with subscripts ``$1$'' (or $\sfrac12$) are already
explicitly known from the solution given by ref.\ \cite{CF}.  The unknown
quantities, with subscripts ``$2$'' (or $\sfrac32$) thus constitute a
relatively manageable part of these complicated looking equations.  Having
thus derived the second order equations, we are now ready to find their 
solutions.

\section{Solutions}

We wish to solve the perturbed Einstein equations
(\ref{00eq}-\ref{55eq}) for the unknown quantities $(\dot a_0/a_0)^2_{(2)}$ and
$(\ddot a_0/a_0)_{(2)}$, which are needed to find the $\rho^2$ corrections to
the Friedmann equations on the TeV brane.
To simplify this task, we will henceforth work in
the limit of stiff
brane potentials, $\lambda_i\to\infty$, which ensures that
$\delta\Phi_i=0$ at each brane.  This allows us to choose a gauge in which
$\delta\Phi_1$ vanishes everywhere in the bulk \cite{CF}, and to solve 
explicitly for $b_1$:\footnote{In fact, the procedure of using a 
small coordinate tranformation 
to set $\delta\Phi_1=0$ can be reiterated to set $\delta\Phi_i(t,y)=0$ to all orders, so that all
effects of $\rho$ and $\rho_*$ on the bulk stress-energy component $T_{55}$ 
are contained in $b_i(t,y)$.}
\beqa
\label{deltab}
{\delta b_1\over b_0} = {b_0\over 2\Phi_0'^2} \left[\Omega^4(\rho-3p) G 
	+ (\rho_*-3p_*)(G-A_0'e^{4A_0})\right]
\eeqa
where 
\beq 
G(y) = \left[\sfrac12
e^{2A_0(y)} + A_0' e^{4A_0(y)} \int_0^y e^{-2A_0} dy \right] / \int_0^1
e^{-2A_0} dy %\cong {kb_0 e^{4kb_0y}\over 1-\Omega^2}, 
\eeq
The other terms which are first order in $\rho$ were also found in
\cite{CF} to be:
\beqa
\label{Psieq}
	\Psi_1 &=& {\kappa^2 b_0\over 6(1-\Omega^2)}\, e^{4 A_0(y)}
	\left(
	F(y) (\Omega^4\rho + \rho_*) - 
	(\Omega^4\rho + \Omega^2\rho_*)\right)\\
\label{Upseq}
	\Upsilon_1 &=& {\kappa^2 b_0\over 2(1-\Omega^2)}\, e^{4
	A_0(y)} \left(
	-F(y) (\Omega^4(\rho+p) + \rho_*+p_*)
%	\right.\nonumber \\ &+&\left.
	+ (\Omega^4(\rho+p) + \Omega^2(\rho_*+p_*))\right)\nonumber\\
\eeqa	
where
\beq
\label{Feq}
	F(y) = 1 - (1-\Omega^2) {\int_0^y e^{-2 A_0} dy \over
	\int_0^1 e^{-2 A_0} dy } %\cong  e^{-2 k b_0 y}
\eeq
and 
\beq
\Omega = e^{-A_0(1)}
\eeq
is the warp factor evaluated at the TeV brane.
  
We recall that the zeroth order static solutions are approximately
(ignoring higher order in $v_0^2$ corrections) \cite{GW2,dewolfe}
\beq
\label{approxsoln}
	\Phi_0(y) \cong v_0 e^{-\eps k b_0 y};\qquad
	A_0(y) \cong kb_0y + \frac{\kappa^2}{12}v_0^2 (e^{-2\eps kb_0y} -1)
\eeq
where we have normalized $A_0(0) = 0$, and introduced
\beq
	\eps = \sqrt{4 + {m^2\over k^2}} - 2  \cong {m^2\over 4 k^2}.
\eeq
Here
\beq 
	k = \sqrt{-\kappa^2\Lambda/6}
\eeq
is the inverse AdS curvature scale,
which should be somewhat below the Planck scale so that higher
derivative operators like $R^2$ do not invalidate the solution.

The above solution is an approximate one in the limit of small $v_0$ for the
model considered by GW \cite{GW2}, where the bulk scalar $\Phi$ has only a mass
term.  It was shown by ref.\ \cite{dewolfe} that this solution is actually {\it
exact} when $\Phi$ has a quartic interaction which is related to the mass in a
certain way.  We are therefore justified in
considering  (\ref{approxsoln}) to be exact, for the appropriate choice
of $V(\Phi)$, in all that will follow.  If on the other hand we decided to
use the pure $m^2\Phi^2$ potential for $V(\Phi)$, the solutions (\ref{approxsoln})
receive corrections which are a power series in $\kappa^2 v_0^2 e^{-2\eps kb_0y}$:
\beqa       
	A_0(y) &=& k b_0 y + {\kappa^2 \phi_0^2\over 12} (e^{-2 k b_0 y}-1)
            +  {\kappa^4 \phi_0^4 \eps\over 96(1+\eps)}(e^{-4k b_0 y}-1) +
	\dots \nonumber\\
        \phi_0(y) &=& \phi_0 e^{-kb_0 y} + {\kappa^2 \phi_0^3 \eps\over 12(1+\eps)}
	 e^{-3 k b_0 y} + \dots
\label{approxsoln2}
\eeqa
Here $\phi_0$ differs from $v_0$ by terms of order $\kappa^2 v_0^3$, to satisfy
the boundary condition $\phi_0(0)=v_0$.
We shall find that none of our main results depend on which bulk scalar potential
is used.

As was previously shown, the first two Einstein equations yield
differential equations for $\Psi_2$ and $\Upsilon_2$.  They are solved by:
\beqa
\Psi_2 &=& e^{4A_0(y)}\left[C_{\Psi} + \int_0^y e^{-4A_0(y)}
\left({\cal F}_{\Psi}(y,t) - \left({\dot a_0\over a_0}\right)_{\!\!(2)}^2 b_0^2
e^{2A_0}\right)dy\right]\\
\Upsilon_2 &=& e^{4A_0(y)}\left[C_{\Upsilon} +
\int_0^y e^{-4A_0(y)} \left( 2\left( \left({\dot a_0\over
a_0}\right)_{\!\!(2)}^2 -
 	\left({\ddot a_0\over a_0}\right)_{\!\!(2)}\right)  b_0^2 e^{2A_0(y)}
	-{\cal F}_{\Upsilon} (y,t)\right) dy\right].\nonumber\\
\eeqa
We do not require the $\Phi$ equation of motion, since this is derivable from
the Einstein equations.  
The boundary conditions (\ref{Psibc},\ref{Upsbc}) allow us to eliminate
the constants of integration $C_{\Psi}$ and $C_{\Upsilon}$ to find
$\left({\ddot a_0/ a_0}\right)_{(2)}$ and
$\left({\dot a_0/a_0}\right)_{(2)}^2$:
\beqa
\label{a2eq1}
\left({\dot a_0\over a_0}\right)_{\!\!(2)}^2 &=& \frac{1}{b_0^2 \int_0^1
e^{-2A_0}dy}\left[\Psi_2(0)-\Omega^4 \Psi_2(1)+\int_0^1 e^{-4A_0(y)} {\cal
F}_{\Psi}dy\right]\\ 
\label{a2eq2}
\!\!\!\!\!\!\!\!\!\! \left({\dot a_0\over a_0}\right)_{\!\!(2)}^2 -
\left({\ddot a_0\over a_0}\right)_{\!\!(2)}&=&\frac{1}{2 b_0^2 \int_0^1
e^{-2A_0}dy}\left[\Omega^4 \Upsilon_2(1) -\Upsilon_2(0) + \int_0^1
e^{-4A_0(y)} {\cal F}_{\Upsilon}\,dy\right].
\eeqa 
Notice that the terms $\Psi_2(0)$, $\Psi_2(1)$, $\Upsilon_2(0)$, $\Upsilon_2(1)$,
are given in terms of $\rho$, $\rho_*$, $p$ and $p_*$ by the jump conditions
(\ref{Psibc}-\ref{Upsbc}).  We will
denote the equations of state for matter on the branes by 
\beq
p=\omega \rho, \qquad p_{*}=\omega_{*} \rho_{*}.
\eeq

To evaluate the integrals of ${\cal F}_\Psi$ and
${\cal F}_\Upsilon$ in (\ref{a2eq1}-\ref{a2eq2}), we need the first order solutions $\Psi_1$, $\Upsilon_1$,
given by (\ref{Psieq}-\ref{Upseq}), as well as $\delta A_1$ and $\delta N_1$.  These can
be obtained from the definitions (\ref{PsiUpseqs}) and the choice of gauge
$\delta\Phi_1=0$, as $\delta A_1(y) = \int_0^y (A_0'{\delta b_1\over b_0} 
+ \Psi_1(y))\, dy$
and $\delta N_1(y) = \delta A_1 +  \int_0^y \Upsilon_1(y)\, dy$.  Moreover, we need
$(\dot a_0/a_0)_{(1/2)}$, given by (\ref{1storderH}), and time derivatives of
the first order perturbations, $\delta\dot A_1$, $\delta\dot N_1$, $\delta\dot b_1$.
To reexpress the latter in terms of $\rho$ and $\rho_*$, we use the following 
relations, which can be derived from the (05) Einstein equation
\beqa
\label{tderivs}\dot \rho &=& -3{\dot a_0 \over
a_0}\!(1+\omega)\,\rho + O(\rho^{5/2})
  \cong - \sqrt{24\pi G(\rho_*+\Omega^4\rho)} (1+\omega)\,\rho \nonumber\\ 
\dot \rho_{*}
 &=& -3{\dot a_0 \over
a_0}\!(1+\omega_{*})\,\rho_{*}+ O(\rho_*^{5/2})
\cong - \sqrt{24\pi G(\rho_*+\Omega^4\rho)} (1+\omega_*)\,\rho_*
\nonumber\\ \ddot \rho &\cong& 12\pi G
(1+\omega)\rho\left[3(1+\omega)\Omega^4 \rho + (3 +\omega_{*}
+2\omega)\rho_{*}\right]\nonumber\\ \ddot \rho_{*} &\cong& 12\pi G
(1+\omega_{*})\rho_{*}\left[3(1+\omega_{*})\rho_{*} + (3 +\omega
+2\omega_{*})\rho \Omega^4\right]
\eeqa
where we have assumed that $\omega$ and $\omega_*$ are constant in time.
(This assumption would only be important during a period of transition
such as going from radiation to matter domination.)
In these expressions, the 4-D effective Newton's constant is defined by integrating
the 5-D gravitational action over the extra dimension to obtain
\beq
\label{8piG}
	8\pi G =\kappa^2\left(2b_0
	\int_0^1e^{-2A_0}dy\right)^{-1}.
\eeq

We now have general expressions allowing us to obtain the second order
corrections to the Friedmann equations.  However the
integrals appearing in these expressions (of the form $\int e^{{\rm
const.}\times A_0}dy$) cannot be performed
analytically for $A_0$ given by (\ref{approxsoln}).  To overcome this
difficulty, we will need to make a further approximation, 
namely that the radion mass is small.  In the small-$v_0$  limit, the radion
mass was found to be
\beq
\label{radmass}
 m_r^2\cong\frac{4}{3}\kappa^2 v_0^2\eps^2 k^2\Omega^{2+2\eps}
\eeq
in ref.\ \cite{CGK}, which also agrees up to factors of order unity with
the result found by ref.\ \cite{GW3}.
By expanding in the small parameter $\eta\equiv\kappa^2 v_0^2/12$, 
we can express the warp factor as 
%\beqa
$e^{c A_0}= e^{ c (k b_0 y + \eta [e^{-2\eps k b_0 y}-1])} %\nonumber\\ &=&
= e^{c k b_0 y} \sum_n \frac{1}{n!} (c\eta[ e^{-2\eps k b_0 y}-1])^n. $ % ;\\
%Z&=&\frac{\kappa^2 v_0^2}{12}.
%\eeqa
The integrals can then be performed exactly, order by order in powers of $v_0^2$. 
In fact, the appearance of $1/\Phi_0'^2$ in (\ref{deltab}) means that the 
leading order term in this expansion will actually be $O(v_0^{-2})$).

At this point we are ready to combine the terms like $(\dot a_0/a_0)^2_{(2)}$ and 
$(\ddot a_0/a_0)_{(2)}$ with those coming from $(\dot a_0/a_0)_{(1/2)}\delta \dot
A_1$  and $\delta \ddot A_1$, to obtain the complete results for $\dot a/a$ and
$\ddot a/a$, as dictated by eqs.\ (\ref{Feq1},\ref{Feq2}).  However, these are not
yet the  physical Hubble rate nor acceleration when evaluated on the TeV brane,
due to a further correction from $\delta N_1$ that must be applied, and which will
be explained in the next section.  Because $\delta A_1$ and $\delta N_1$  depend
on $y$, the rates $\dot a/a$ as measured by observers on the two branes will
differ.

\section{Physical Friedmann Equations on the Branes}

Although the main results found above, (\ref{a2eq1})-(\ref{a2eq2}),  will resemble
the Friedmann equations when expressed as functions of the  energy densities on
the branes (after being combined with the appropriate corrections from $\delta\dot
A_1$ and $\delta\ddot A_1$), they are not yet  written in terms of the  standard
Friedmann-Robertson-Walker (FRW) time variable for TeV brane observers, which we
shall denote by $\tau$.  On the TeV brane, the lapse function $n^2(t,1)$ has also
received corrections of the kind we are interested in.  Therefore we should
transform to $\tau$ using
\beq
\label{taueq}
	d\tau = {n(t,1)\over n_0(t,1)}\, dt = e^{-\delta N_1(t,1) -
	\delta N_2(t,1) - \cdots}\, dt
\eeq
where $n_0(t,1) = a_0(t,1) = \Omega$ is the warp factor without any
perturbations due to matter.  

At first it may not be obvious why we should use only the perturbations,
$\delta N_i$, to relate $\tau$ to $t$ instead of using the full warp
factor.  Of course, one is always free to rescale the time coordinate by a
constant factor; this cannot change any physical observables.  There are
several ways of seeing why the above choice is the most straightforward
one.  First, it gives the same time coordinate as that in which the
resolution of the gauge hierarchy problem was couched in the original RS
paper \cite{RSI}.  Specifically, using the time coordinate as defined
above, in the absence of the cosmological perturbation, the action for
a free scalar field on the TeV brane has the form
\beqa
	S &=& \sfrac12 \int d^{\,4}\!x\, \Omega^4
\left(\Omega^{-2}\left((\dot\phi^2) - (\nabla\phi)^2\right) - m_0^2\phi^2
\right)\nonumber\\
	&=&  \sfrac12 \int d^{\,4}\!x\,
	\left(\left((\dot\varphi^2) - (\nabla\varphi)^2\right) - \Omega^2 
	m_0^2\varphi^2   \right)
\eeqa
where $\varphi = \Omega\phi$.  This was the argument used to show
that a bare mass of $m_0\sim M_p$ translates into a physical mass of 
$m = \Omega m_0$.  If on the other hand we had defined $d\tau = \Omega dt$
as we might have been tempted to do in (\ref{taueq}), we would get
\beqa
        S &\to& \sfrac12 \int d\tau\,d^{\,3}\!x\, \Omega^3
\left((\dot\phi^2) - \Omega^{-2} (\nabla\phi)^2 - m_0^2\phi^2
\right)\nonumber\\
	&=& \sfrac12 \int d\tau\,d^{\,3}\!x 
\left( (\dot\varphi^2) - \Omega^{-2}(\nabla\varphi)^2 - m_0^2\varphi^2
\right)
\eeqa
where now $\varphi = \Omega^{3/2}\phi$.  In these variables it is no
longer obvious that physical masses are suppressed relative to the Planck
scale.  The resolution is that one must also reconsider the 4-D effective
gravitational action in the new time coordinate.  The result is that the
4-D Planck scale is exponentially larger than the underlying 5-D gravity
scale, $(8\pi G)^{-1} = \Omega^{-2} \kappa^{-2} / k$.  In this picture,
the fundamental scales are all taken to be of order TeV, while the 4-D
Planck scale is enhanced by $\Omega^{-1}$.  Another, simpler, way of
deriving the same result is to do everything with the warp factor defined
to be unity on the TeV brane, and large on the Planck brane.  In either
method, the ratio of physical particle masses to the 4-D Planck mass
is the same.

The consequence of all this is simple: to find the physical Hubble rate 
at some position in the bulk $y$ in
terms of the complete expression for $\dot a/a$, one merely multiplies by
$dt/d\tau = e^{\delta N_1 + \delta N_2}|_y$:
\beq
\label{dNcorr}
	H \cong (1 + \delta N_1)
	{\dot a\over a} 
\eeq
Since the leading contribution to $H$ is already $O(\rho^{1/2})$, we can
ignore the $\delta N_2$ corrections; these would be
$O(\rho^{5/2})$.  Similarly, the second Friedmann equation, which goes
like $\dot H$, becomes
\beqa
   {d H\over d\tau} &=& (1 + \delta N_1) {d\over dt} H	
	\nonumber\\
	&\cong & (1 + 2 \delta N_1)\left( {\ddot a\over a} - \left({\dot
a\over a}\right)^2 \right) + \delta\dot N_1 {\dot a_0\over a_0}
\eeqa
to the order we are calculating.  All dots are still time derivatives with
respect to $t$, not $\tau$.

The general expressions for the order $\rho^2$ corrections to $H^2$ and $dH/d\tau$
are rather cumbersome if expressed at an arbitrary position in the bulk.  Here we
give the results for $H$ evaluated on either the TeV or the Planck brane, to
leading order in the radion mass squared, $m^2_r$, eq.\ (\ref{radmass}).  We have defined $\bar\rho =
\Omega^4\rho$ since  this is the physically observable energy density on the TeV
brane, whereas $\rho$ is the bare value.  For each term $\rho_*^2$,
$\rho_*\bar\rho$, $\bar\rho^2$, we give only the leading dependence on the warp
factor $\Omega$, which is usually assumed to be $\ll 1$ to solve the hierarchy
 problem.  Thus we have found the leading corrections to the Hubble rate and
acceleration in a simultaneous
expansion in $\rho$, $\rho_*$, $1/m_r$ and $\Omega$:\footnote{We have factored
these expressions to make it more clear that they vanish when
$\omega=\omega_*=1/3$.  The asymmetrical appearance of terms like
$2(1-3\omega)(2+3\omega) - 3(1-3\omega_*)(1+\omega)$ is not due to a 
typographical error.}
\beqa
%%%%%%%%%%%%%%%%%%%%%
\label{H21}
H^2|_{y=1} &=& {8\pi G\over 3} \left( \bar\rho + \rho_* +
	{2\pi G\over 3 m^2_r\Omega^2} \left( 
9(1-3\omega)(1+\omega)\bar\rho^2  \right.\right.\nonumber\\
 && \left.\phantom{8\pi G\over 3}\left. +
4(1-3\omega_*)(4+3\omega_*)\Omega^2\rho_*^2
+4(1-3\omega)(4+3\omega)\bar\rho\rho_* ) \right)\right)\\
%%%%%%%%%%%%%%%%%%
\label{H20}
H^2|_{y=0} &=&   {8\pi G\over 3} \left( \bar\rho + \rho_* +
	{2\pi G \over 3 m^2_r\Omega^2} \left( 
9(1-3\omega_*)(1+\omega_*)\rho_*^2\Omega^4  -
(1-3\omega)(7+3\omega)\bar\rho^2
\right.\right. \nonumber\\
 && \left.\phantom{8\pi G\over 3}\left. 
+ 2\Omega^{2}\bar\rho\rho_*[ 2(1-3\omega)(2+3\omega) - 3(1-3\omega_*)(1+\omega) 
 ] \ \right)\right)\\
%%%%%%%%%%%%%%%%%%%
\label{dH1}
\left.{dH\over d\tau}\right|_{y=1} &=&-4 \pi G \left(\bar\rho (1+\omega) 
+ \rho_* (1+\omega_*) \phantom{1\over 3}\right.\nonumber\\
  &+& \left.\phantom{-4 \pi G\!\!\!\!\!\!\!\!\!\!\!\!\!\!\!\!}
{4 \pi G \over 3 m_r^2 \Omega^2}
\left(\Omega^2(1+\omega_*)(1-3\omega_*)(13+9\omega_*)
\rho_*^2+9(1-3\omega)(1+\omega)^2\bar\rho^2
\phantom{1\over 3}\!\!\!\!\!\right. \right. \nonumber\\
 &-&\left.\left. \phantom{-4 \pi G\!\!\!\!\!\!\!\!\!\!\!\!\!\!\!\!}
(1-3\omega)(2(1+\omega_*)
+2(1+\omega)+6(1+\omega)^2+3(1+\omega)(1+\omega_*))
{\bar\rho}\rho_* \right)\phantom{1\over 3}\!\!\!\!\!
\right)\nonumber\\
\\
%%%%%%%%%%%%%%%%%%%
\label{dH0}
\left.{dH\over d\tau}\right|_{y=0} &=&-4 \pi G 
\left(\bar\rho (1+\omega) + \rho_* (1+\omega_*)\phantom{1\over 3} \right.\nonumber\\
 &+&\left. \phantom{-4 \pi G\!\!\!\!\!\!\!\!\!\!\!\!\!\!\!\!} 
{4 \pi G \over 3 m_r^2 \Omega^2}\left(-4(1+\omega)
(1-3\omega)\bar\rho^2+9(1+\omega_*)^2(1-3\omega_*)\Omega^4 \rho_*^2
\phantom{1\over 3}\!\!\!\!\!\right.\right. \nonumber\\
&+&\left.\left.\phantom{-4 \pi G\!\!\!\!\!\!\!\!\!\!\!\!\!\!\!\!}
\Omega^2\left[6(1-3\omega)(1+\omega)^2+2(1-3\omega)(1+\omega)-2(1-3\omega)
(1+\omega_*)\right.\right.\right.\nonumber\\
&-&\left.\left.\left.\phantom{-4 \pi G\!\!\!\!\!\!\!\!\!\!\!\!\!\!\!\!} 
4(1-3\omega_*)(1+\omega)+3(1+\omega_*)(1-3\omega)(1+\omega)
\phantom{\Omega^2\!\!\!\!\!\!\!}\right]
\rho_* {\bar\rho}\phantom{1\over 3}\!\!\!\right)\right)
\eeqa
These, finally, can be considered to be the main results of this paper, except in
the case of $\omega=\omega_*=1/3$.  As can be seen, in that case
the quadratic corrections vanish
at order $1/m^2_r$, so we have to go to higher order in $m^2_r$ to get the
leading result for radiation, as will be described
below.   Two comments are in order.  (1) These results are independent of
the choice of bulk potential $V(\Phi)$ for the scalar.  Any dependence on $V$
comes in starting at the next order in $v_0^2$.  (2) It may seem strange that
on the Planck brane ($y=0$) powers of $\Omega^2$ track powers of $\rho_*$,
but not so on the TeV brane.  The difference is due to the corrections to
$\dot a_0\over a_0$ from $\delta \dot A_1\sim H\delta A_1$, eq.\ (\ref{Feq1}), 
and $\delta N_1$, eq.\ (\ref{dNcorr}).  They give contributions to $H^2$ of
order $H^2 \times[\delta A_1(1),\delta N_1(1)]$.  Since $H^2\sim \rho_*+\bar\rho$
at lowest order, the correlation of powers of $\Omega^2$ and powers of $\rho_*$
is no longer respected.  

Equipped with these results, we can now specialize to several situations
of cosmological interest.

\subsection{Inflation}

The earliest epoch of interest is an inflationary one, where the equation of state
is $\omega=-1$ or $\omega_*=-1$, depending on whether we put the  inflaton on the
TeV or the Planck brane.  Let us first suppose the inflaton is on the the TeV 
brane, so that $\rho_*=0$.  Then eq.\ (\ref{H21}) tells us that there is no
quadratic correction to the Hubble rate at $y=1$, in contrast to the situation in
pure RS-II (single brane) cosmology.  Not only is this true for the leading term
in the expansion in $m_r$, but in fact we have checked that it is true order by
order for the terms $1/m_r^2$ and $m_r^0$.  The fact that many terms must 
seemingly miraculously cancel to achieve this suggests that it is true to all
orders.

The analogous statements hold true for an inflaton on the Planck brane: if
$\bar\rho=0$ and only $\rho_*$ is nonzero, there is no quadratic correction to
$H^2$ at $y=0$.   Again, we have shown this to be exactly true for the first few
orders in an expansion in $m^2_r$. There {\it is} such a correction to $H^2$ at
the TeV brane, but unfortunately this has no impact on the rolling of a scalar
field on the Planck brane, since its own Hubble rate is unaffected.  Thus we
cannot take advantage of the modification to the Friedmann equation to admit
steeper inflationary potentials than are normally allowed by the slow-roll
conditions, as was explored in refs.\ \cite{MWBH,CLL}.   Only if both $\rho_*$ and
$\bar\rho$ are nonzero do we get this kind of effect (through the $\rho_*\bar\rho$
term in eq.\ (\ref{H20})).  But even there, its sign tends to be the wrong one for
the desired purpose.  For example, suppose that some of the inflaton's energy
density has been converted energy on the TeV brane, with an arbitrary value of
$\omega$ while $\omega_*=-1$.  Then (\ref{H20}) predicts that $H^2$ is decreased
by the quadratic corrections, regardless of the value of $\omega$.

Even though there might be no presently observable effect,  it is still of
theoretical interest to consider the modification to the TeV brane's Hubble rate
due to a Planck brane inflaton, to contrast with simpler models of brane
cosmology.  With $\rho=0$ but $\rho_*\neq 0$ and $H^2|_{y=1}$ is corrected by the
factor
\beq
  H^2|_{y=1} = {8\pi G\over 3}\rho_*\left(1 + {32\pi G\over 3m^2_r}\rho_*
\right)
\eeq
The correction becomes important at temperatures of order the intermediate scale
$T \sim \sqrt{m_r M_p} \sim 10^{10}$ GeV.  We can compare this to the situation
where there is no stabilizing scalar field, but a fine-tuning $\rho =
-\rho_*/\Omega^2$ is required between the energy densities on the two branes, in
order to keep the bulk from expanding.  In this case, the relation (\ref{mfe})
applies to the given order at both branes, since for $\omega_*=-1$ the difference
$H^2(0)-H^2(1) = O(\rho_*^3)$ \cite{CGRT}.  But this correction only becomes
important when $T\sim M_p$ in the unstabilized case, since the natural scale for
the Planck brane tension is $M_p^4$.  

The effect in the present case is easy to understand in terms of the usual slow
roll condition for scalar fields during inflation.  If the Hubble rate exceeds the
mass of the radion, it will start rolling, hence the size of the extra dimension
is destabilized, which in turn changes the strength of gravity from the 4-D point
of view, due to the dependence of $G$ on $b_0$ in (\ref{8piG}).  The criterion
that $H<m_r$ translates to $T<\sqrt{m_r M_p}$, which is precisely the condition we
found for the $\rho_*^2$ correction to $H^2$ to be small.  On the other hand, the
reader should not be misled into thinking that the $\rho_*^2$ term can be deduced
from this effect (Hubble expansion giving an $m^2_r$-dependent shift in $b$,
leading to a shift in $G$ alone).  The coefficient of $\rho_*^2$ is numerically
different from that which is obtained solely from the effect of shifting $G$.

We noted that an inflaton on the TeV brane gives no $\bar\rho^2$ correction to
$H^2$ measured on the TeV brane.   A similar story applies for an observer on the
Planck brane: he sees no $\rho_*^2$ correction to $H^2$ on his brane.  Again, we
find this to be true order by order in an expansion about $m^2_r=0$, so it is
presumably an exact statement.  Moreover we have found this to be true for either
of the choices of the bulk potential $V(\Phi)$ that we have considered. At first
sight this may be a surprising result, because it holds even in the limit that the
infrared brane is removed to $y\to\infty$.  In this limit we might expect physics
to coincide with that of the RS-II model, where the $\rho_*^2$ correction to
$H^2|_{y=0}$ (eq.\ (\ref{mfe})) is well known to occur, at least in the case of no
bulk scalar field.  We will explain the apparent discrepancy in a subsection
below.

\subsection{Radiation era}

Our results (\ref{H20}-\ref{dH0}) predict that there is no effect of order 
$\rho^2/m^2_r$ when the equation of state on the two branes is that of radiation,
$\omega=\omega_*=1/3$.  The origin of the 
$1/m^2_r$ dependence is the large shift in $\delta b_1$
(\ref{deltab}) for generic equations of state.  In the special case of radiation,
$\delta b_1=0$ since the radion couples to the trace of the stress energy tensor,
which vanishes in this case.
We have therefore done a separate treatment when
$\omega=\omega_*=1/3$ to find the leading effect.  We obtain
\beqa
%%%%%%%%%%%%%%%%%%%%%
  H^2|_{y=1} &=& {8\pi G\over 3}\left((\rho_*+\bar\rho) +
\frac{2 \pi G}{3k^2 \Omega^4} % (1-\Omega^2)^2
	(\Omega^2\rho_*-\bar\rho)\rho_*\right)\\
%%%%%%%%%%%%%%%%%%%%%
H^2|_{y=0} &=& {8\pi G\over 3}\left((\rho_*+\bar\rho) -
\frac{2 \pi G}{3k^2 \Omega^6} % (1-\Omega^2)^2
	(\Omega^2\rho_*-\bar\rho)\bar\rho\right)
\eeqa
and the equations for $dH/d\tau$ have vanishing corrections at this order in
$\rho$ and $\rho_*$.  It is curious that, 
just like in the previously discussed case of inflation,
the Hubble rate on a given brane receives an order $\rho^2$ correction only if 
the energy density on the {\it other} brane is nonvanishing.  

We can make the preceding results much stronger in the case where the
energy density vanishes on one of the two branes.
If $\rho_* = 0$, then eq.\ (\ref{a2eq1}) implies that 
the second order correction to the
Hubble rate on the TeV brane vanishes identically, with no approximations,
provided that
\beqa
&&\int_0^1\left[e^{4A_0(y_1)}\left(\int_0^{y_1} e^{-2A_0(y_2)}dy_2\right)
\left(\int_0^1 e^{-2A_0(y_3)}dy_3\right)-e^{4A_0(y_1)}
\left(\int_0^{y_1} e^{-2A_0(y_2)}dy_2\right)^2\right.\nonumber\\
&&\qquad\qquad -\left.e^{-2A_0(y_1)}\int^{y_1}_0\left(e^{4A_0(y_2)}
 \int_0^{y_2} e^{-2A_0(y_3)}dy_3\right)dy_2
\right]dy_1 =0
\eeqa
In the inverse case where $\rho = 0$ and $\omega_* = 1/3$, the second order 
correction to $H^2$ evaluated on the Planck brane vanishes if
\beqa
&&\int^1_0\left[e^{4A_0(y_1)}\left(\int^{y_1}_0 e^{-2A_0(y_2)}dy_2 
- \int^1_0 e^{-2A_0(y_2)}dy_2 \right)^2 \right.\\
&&\qquad+ \left.e^{-2A_0(y_1)}\int^{y_1}_0 e^{4A_0(y_2)}\left(\int^{y_2}_0 
e^{-2A_0(y_3)}dy_3 - \int^1_0 e^{-2A_0(y_3)}dy_3 \right)dy_2\right]dy_1=0\nonumber
\eeqa
Writing
\beqa
u_1 &=& \int^{y_1}_0\left(e^{4A_0(y_2)} \int_0^{y_2} e^{-2A_0(y_3)}dy_3
\right)dy_2\nonumber\\
u_2 &=& \int^{y_1}_0 e^{4A_0(y_2)} \left(\int_0^{y_2} e^{-2A_0(y_3)}dy_3 -
 \int_0^1 e^{-2A_0(y_3)}dy_3 \right)dy_2\nonumber\\
dv &=& e^{-2A_0(y_1)}dy_1 
\eeqa
where $u_1$ and $u_2$ are to be used in the first and second case respectively, 
we can use integration by 
parts ($\int u_i dv = u_i v - \int v du_i$) 
to show that these integrals do indeed vanish identically.  

Let us now focus on physics on the TeV brane $(y=1)$.  By the time of 
nucleosynthesis
no more than 10\% of the radiation energy density can be on the Planck brane,
so one expects that $\rho_* \lsim \bar\rho$ at TeV temperatures as well.
Therefore we can ignore the $\Omega^2\rho_*$ term, and the fractional correction
to $H^2$ is of order
\beq
	{\delta H^2\over H^2}-1 \sim - \left({\rho_*\over {\rm TeV}^4}\right)
\eeq
It is interesting that the order of magnitude is nearly correct for this to be
relevant at the time of the electroweak phase transition.  (We could change TeV
$\to$ 100 GeV by making $k$ a few orders of magnitude smaller than $M_p$.) However
the minus sign shows that the Hubble rate is suppressed relative to  standard
cosmology.  This is disappointing for electroweak baryogenesis, since one would
have liked to increase $H^2$ make sphalerons go out of equilibrium at the
electroweak phase transition, a possibility that has been recently analyzed in
ref.\ \cite{Geraldine}. Of course, it may happen that the sign turns out to be the
right one in some other model of brane cosmology.

\subsection{Relation to RS-I model without stabilization and to RS-II}

As a check on our results, one would like verify that they are
consistent with related cosmological solutions.  For example, when there
is no scalar field and the equation of state is inflationary,
Kaloper \cite{Kaloper} found the exact solution\footnote{This is related to
the coordinate system we have been using by $w = by$, where now $b$ is taken to
be constant by fiat.}
\beqa
\label{Kaloper}
	ds^2 &=&  a^2(w) \left(
	-dt^2 + e^{2Ht} d\vec x^{\,2} \right) + dw^2;\nonumber\\	
	a(w) &=& e^{-k w} - \hat\rho_*\sinh k w
\eeqa
where $\hat\rho_* = \rho_*/{\cal T}$, the ratio of the excess energy density
to the tension on the Planck brane, and the Hubble rate is given by
\beq
\label{mfe2}
	H^2 = 2 k^2((1+\hat\rho_*)^2-1)
\eeq
As in the static case, one can decide whether or not to compactify the extra
dimension by inserting a second brane at some position $w_1$.  Without 
compactification, the extra dimension is still effectively cut off by the
presence of a horizon at $w={1\over 2k}\ln(1 + {2\over \hat\rho_*})$.
If we compactify explicitly by inserting a second brane, its 
tension ($-{\cal T}$) 
and excess energy density $\rho$ are completely fixed by the value of $a'(w_1)$, 
leading to the constraint
\beq
	\hat\rho\equiv {\rho\over {\cal T}}  = -1 + 
	{e^{-kw_1} -\hat\rho_* \cosh kw_1 \over e^{-kw_1} -\hat\rho_* \sinh kw_1}
\eeq
Expanding in $\rho_*$, to leading order this gives the well-known constraint
\cite{CGRT} $\rho = -\rho_*/\Omega^2$ needed for a static bulk in the absence
of stabilization (recall that the $\Omega=e^{-kw_1}$ is the warp factor).  However, the exact tuning is a power series in $\rho_*$, and
our second order formalism correctly reproduces the $O(\rho_*^2)$ term, as we
describe in more detail in appendix C.  Of course, our perturbative treatment
also correctly reproduces the solution $a(w)$ and $a_0(t)$ expanded in powers 
of $\rho_*$.

The previous check was specifically for the equation of state $\omega_*=-1$.
The same can be done for an arbitrary equation of
state, for which the (\cite{Kaloper}) solution generalizes to \cite{BDEL,MSM}
\beqa
\label{genform}
	ds^2 &=&  - n^2(t,w) dt^2 + a^2(t,w) d\vec x^{\,2} + dw^2;\nonumber\\
	a(t,w) &=& a_0(t)( e^{-k w} - \hat\rho_*(t)\sinh k w);\nonumber\\
	n(t,w) &=& e^{-k w} + \hat\rho_*(t)(2+3\omega_*)\sinh k w ,
\eeqa
where $a_0(t)$ is determined by the modified Friedmann equation ({\ref{mfe2})
with $H={\dot a_0\over a_0}$.
Again, our perturbative formalism agrees with the expansion of the exact
solution at order $\rho_*^2$.   It is interesting to note in passing that
for $\omega_*>-1$, the metric function $a(t,y)$ vanishes at a position between
the brane at $y=0$ and the horizon where $n(t,y)=0$.  Nevertheless, all components
of the 5D Riemann tensor are well behaved throughout the bulk, and neither the
vanishing of $a(t,y)$ nor $n(t,y)$ leads to any curvature singularities.

Now we come to a slightly more subtle issue: what happens if we try to smoothly
remove the TeV brane from the problem, to recover the known single-brane results? 
We have noted that our results for $H^2$ do not reduce to those of the RS-II
model, even in the case where the TeV brane is taken to infinity. The presence of
the bulk scalar is crucial for understanding the difference between the two
situations, because it modifies the bulk stress-energy component $T_{55}$ in
response to matter on the branes.  Even though we are working in a gauge where
$\delta\Phi_i=0$, $T_{55} = g_{55}(V(\Phi)+\Lambda) -\frac12\Phi'^2$ is
nevertheless modified by the shift in $g_{55} = b(t,y)^2$.  It is the $G_{55}=
\kappa^2 T_{55}$ Einstein equation which implies the exotic form of the Friedmann
equation $H^2\sim (\rho_*+{\cal T})^2$ when there is no matter in the bulk.  This
argument no longer holds when $T_{55}$ is present. Therefore it is not so
surprising that the $\rho_*^2$ corrections no longer coincide with those of the
RS-II model, when we have just a single brane but also a scalar field.  

Mathematically, we cannot recover the RS-II model simply by taking the limit
$v_0\to 0$, which corresponds to removing the scalar field from the problem.  If
we attempt  to do this, our perturbation expansion breaks down because $\delta
b_1$, eq.\ (\ref{deltab}), diverges---{\it unless} the matter on the branes is
tuned so that the term in brackets in (\ref{deltab}) vanishes.  This tuning is 
the condition $\rho = -\rho_*/\Omega^2$ discussed above.  A special case is that
of radiation, where no such breakdown of perturbation theory occurs.

\subsection{Dark Radiation}

In the present work we have neglected the presence of dark radiation
\cite{BDEL,kraus,ida,dark-rad}, which appears in the Friedmann equation in the normal way,
\beq
	H^2 = 2 k^2\left( (\hat\rho_*+1)^2-1 + {c\over a_0(t)^4}\right);
\eeq
in particular, there is no quadratic correction ${c^2/a_0(t)^8}$ even
in the unstabilized case.  The AdS/CFT interpretation is that the dark radiation
is simply the thermalized degrees of freedom of the conformal field theory
\cite{Gubser}, but
from the 5D viewpoint, the dark radiation is associated with the presence of a
black-hole-like singularity that appears in the bulk at a finite distance from
the brane.  This can be most clearly seen in the coordinate system of ref.\ 
\cite{kraus} (see also \cite{BCG}), where the brane is moving through a static bulk whose geometry is
that of the 5D AdS-Schwarzschild metric.

Again using the results of \cite{BDEL}, the exact
solution takes the form (\ref{genform}), but now
\beqa
	a^2(t,w) &=& a_0^2(t)\left( e^{-k w} - \hat\rho_*(t)\sinh k w\right)^2
	+ {c\over a_0(t)^2}\sinh^2 k w; \nonumber\\
	n(t,w) &=& {\dot a(t,w)\over \dot a_0(t)} = {a_0(t)\over
        a(t,y)} \left( \left( e^{-k w} - \hat\rho_*(t)\sinh k w\right)^2\right.
	\\
	&& \left. + 3\rho_*(t)(1+\omega_*) \left( e^{-k w} - \hat\rho_*(t)\sinh k w\right)
	\sinh kw - {2 c\over a_0^4(t)} \sinh^2 kw \right) \nonumber
\eeqa
From this exact solution we see that the metric functions do not factorize
in the way which we assumed in our original ansatz (\ref{ansatz}).  A more
general form would therefore be needed to study the interactions, if any,
between $\rho^2$ corrections and dark radiation in the stabilized model.

\section{Summary and conclusions}

In this paper we have focused on the corrections to the Friedmann equation which
arise in the possibly more realistic 5D brane cosmologies where the bulk  stress
energy is not trivial, but has some inhomogeneity in the extra dimension due to a
scalar field.  The presence of the scalar was motivated by the need to stabilize
the size of the extra dimension when the extra dimension is compactified by the
presence of a second (TeV) brane, but it could also be present on more general
grounds.

In a perturbative expansion in powers of energy densities on the branes, the
Hubble parameter $H$ generally should receive corrections to all orders in
$\rho$ and $\rho_*$ rather than terminating at order $\rho_*^2$ as happens in the
unstabilized solutions.   We found the rather surprising result that the 
$\rho^2$ correction vanishes in the special cases where the energy density
is confined to the same brane as that where $H$ is measured and the equation
of state is either that of inflation or radiation.  If either of these conditions
are not fulfilled, the $\rho^2$ correction does not vanish, but its sign and
magnitude can be different from the case of a single brane with no bulk scalar
field.  In these cases, the correction has a coefficient of order 
(TeV)$^{-2} m^{-2}_r$, where $m_r$ is the mass of the radion, which is expected
to be somewhat lighter than the TeV scale.

We noted two implications of this result.  First, it is impossible to obtain order
$\rho^2$ corrections to $H^2$ during inflation on a given brane unless the
inflaton is on the {\it opposite} brane.  This means the corrections have no
effect on the dynamics of the inflaton, which is disappointing from the  point of
view of potentially observing the effects of branes on cosmology. Second, it is
equally true during radiation domination that $\rho^2$ effects can appear for a
given brane only if the energy density on the other brane is nonzero.  If some
fraction of the radiation of  the universe is on the Planck brane, there could be
significant deviations from standard cosmology during the electroweak epoch. 
Unfortunately, the sign of the corrections is such as to decrease the Hubble
rate.  It would have been more interesting for the purposes of electroweak
baryogenesis to obtain the other sign.

On a more optimistic note, we should reiterate that the exact cosmological
behavior cannot be predicted from the $\rho^2$ terms when they become as important
as important as the linear in $\rho$ terms, since the perturbative expansion is
breaking down.  It would therefore be interesting to find
exact cosmological solutions in the presence of the stabilizing field.

\bigskip
We thank Hassan Firouzjahi for enlightening discussions, and G\'eraldine Servant
for helpful comments on the manuscript.

\appendix
\section{1st order functions appearing in 2nd order Einstein equations} 
These are the functions appearing in eqs.\ (\ref{00eq}-\ref{scalar}), which
give the solutions for the second order perturbations.

\beqa
\label{fPsi}%%%%%%%%%%%%%%%%%%%%%%%%%%%%%%%%%%%%%%%%%%%%%%%%%%%%%%
{\cal F}_{\Psi} = &2&\left(\delta N_1 + {\delta b_1\over b_0}\right)
	\Psi_1' +2\Psi_1^2 - \frac{\kappa^2}{6} \Phi_0' \delta \Phi_1' 
	{\delta b_1\over b_0} + \Psi_1\left({\delta b_1'\over b_0} - 4A_0'\left(2\delta
N_1 + {\delta b_1\over b_0}\right)\right) \nonumber\\
&+& \frac{\kappa^2}{6} \Phi_0'^2 
	\left({\delta b_1\over b_0}\right)^2  + \frac{\kappa^2}{6}\left(\Phi_0' (4 \Psi_1 - \Upsilon_1) +
\Phi_0''{\delta b_1\over b_0} \right) \delta \Phi_1 \nonumber\\
&+& \left({\dot a_0\over a_0}\right)_{\!\!\!\left(\sfrac12\right)}e^{2A_0}
b_0^2
\left({2\delta \dot A_1 -\dot {\delta b_1\over b_0} }\right)\\
\label{fUpsilon}%%%%%%%%%%%%%%%%%%%%%%%%%%%%%%%%%%%%%%%%%%%%%%%%%%%%%%
{\cal F}_{\Upsilon} = &-& 4 \Upsilon_1\Psi_1 - \Upsilon_1^2 - 
   2\left(\delta N_1 + {\delta b_1\over b_0}\right)\Upsilon_1'+ \left(4 A_0'\left(2\delta N_1 + {\delta b_1\over
	b_0}\right)-{\delta
	b_1\over b_0}' \right)\Upsilon_1\nonumber\\
&-&4\frac{\kappa^2}{3}\Phi_0' 
	\Upsilon_1 \delta \Phi_1+ \left(\left({\dot a_0\over
a_0}\right)_{\!\!\!\left(\sfrac12\right)} \left(2\delta {\dot
N_1}-{\dot{\delta b_1\over b_0} }\right) + 
	{\ddot {\delta b_1\over b_0}}-2\delta \ddot A_1\right)e^{2A_0} b_0^2\\
\label{f05}%%%%%%%%%%%%%%%%%%%%%%%%%%%%%%%%%%%%%%%%%%%%%%%%%%%%%%
{\cal F}_{05} = &&\left(\delta
\dot A_1 - \left({\dot a_0\over a_0}\right)_{\!\!\!\left(\sfrac32\right)}
\right)\Upsilon_1 - 
	{\dot{\delta b_1\over b_0} }\Psi_1\nonumber\\
&&+ \frac{\kappa^2}{6}\left( \left(\delta {\dot \Phi_1'} -
\Phi_0'{\dot{\delta b_1\over b_0}} \right)\delta\Phi_1 + 
\left(\Phi_0'{\delta b_1\over b_0}-\delta \Phi_1'\right)
\delta {\dot\Phi_1}\right) \\
\label{f55}%%%%%%%%%%%%%%%%%%%%%%%%%%%%%%%%%%%%%%%%%%%%%%%%%%%%%%
{\cal F}_{55} = &-&\left(2\delta N_1 +{\delta b_1\over b_0} \right)A_0'\Upsilon_1
	+2\Psi_1^2+\left(\Upsilon_1 - 4 A_0'\left(2\delta N_1 +{\delta
b_1\over
b_0}\right)\right)\Psi_1\nonumber\\
&& + \frac{\kappa^2}{3}\left(- 2\Phi_0'^2\delta N_1 
{\delta b_1\over b_0} -\frac{3}{2} \Phi_0'^2 
\left({\delta b_1\over b_0}\right)^2 \right) - \left({\dot a_0\over
a_0}\right)_{\!\!\!\left(\sfrac12\right)}e^{2A_0} b_0^2 \left(\delta {\dot
N_1}-4\delta {\dot A}_1\right) \nonumber\\
&&+ e^{2A_0} b_0^2 \delta {\ddot A}_1
-\frac{\kappa^2}{6} \delta \Phi_1'^2+ 2\frac{\kappa^2}{3}\left(\delta N_1+ {\delta b_1\over
b_0}\right)\Phi_0'\delta\Phi_1'
-\frac{\kappa^2}{6}\delta\Phi_1'^2\nonumber\\
&&+\frac{\kappa^2}{6}\left((\Upsilon_1-{\delta b_1\over
b_0}'+4\Psi_1)\Phi_0'+(-2{\delta b_1\over
b_0}-4\delta N_1)\Phi_0''+\delta\Phi_1''\right)\delta\Phi_1\\
\label{fPhi}%%%%%%%%%%%%%%%%%%%%%%%%%%%%%%%%%%%%%%%%%%%%%%%%%%%%%%
{\cal F}_{\Phi} = &&{\delta b_1\over
b_0}^2\left(-3\Phi_0''+4A_0'\Phi_0'\right)-\Phi_0' {\delta b_1\over
b_0}(8\Psi_1+3{\delta b_1\over b_0}'+2\Upsilon_1)\nonumber\\
&&+\frac{V'''(\Phi_0)}{2}b_0^2 \delta\Phi_1^2 + 2\kappa^2 \Phi_0'
\delta\Phi_1 \left(-{\delta b_1\over b_0}\Phi_0'+\delta
\Phi_1'\right)\nonumber\\ &&+\delta\Phi_1'\left(-4 {\delta b_1\over b_0}
A_0'+\Upsilon_1+4\Psi_1+ {\delta b_1\over b_0}'\right)+2\delta\Phi_1''{\delta
b_1\over b_0}\nonumber\\ &&+b_0^2 e^{2A_0}\left(3 \left({\dot a_0\over
a_0}\right)_{\!\!\!\left(\sfrac12\right)}  \delta {\dot \Phi}_1+\delta{\ddot
\Phi}_1\right)
\eeqa

\section{Complete second order corrections}
In the previous sections, we presented simplified expression for $H^2$ and $\dot H$.  Here we present the complete expressions for the second order in $\rho$ corrections to $O(v_0^{-2})$, evaluated on the branes.

\beqa
%%%%%%%%%%%%%%%%%%%%%
H^2_2 (y=1) = &&\frac{8\pi^2 G^2 (\Omega^{-2\eps}-\Omega^4)}{3 k^2 \kappa^2 v_0^2 \eps^2 (2+\eps)(1-\Omega^2)\Omega^2}\left[9\Omega^6(1-3\omega)(1+\omega)\rho^2 \right.\nonumber\\
&&+ (1-3\omega_*)(4(4+3\omega_*)-\Omega^2(7+3\omega_*))\rho_*^2\nonumber\\
&&\left.+2\Omega^2\left\{\Omega^2((1-3\omega)(1-3\omega_*)+6(1+\omega_*)(1-3\omega_*)-2(1-3\omega_*)-4(1-3\omega))\right.\right.\nonumber\\
&&\left.\left.+2(1-3\omega)+6(1+\omega)(1-3\omega)\right\}\rho\rho_*\right]\\
%%%%%%%%%%%%%%%%%%
H^2_2 (y=0) =  &&\frac{8\pi^2 G^2 (\Omega^{-2\eps}-\Omega^4)}{3 k^2 \kappa^2 v_0^2 \eps^2 (2+\eps)(1-\Omega^2)}\left[9(1-3\omega_*)(1+\omega_*)\rho_*^2 \right.\nonumber\\
&&+ \Omega^4(1-3\omega)(4\Omega^2(4+3\omega)-(7+3\omega))\rho^2\nonumber\\
&&\left.+2\Omega^2\left\{(1-3\omega)(1-3\omega_*)+6(1+\omega)(1-3\omega)-2(1-3\omega)-4(1-3\omega_*)\right.\right.\nonumber\\
&&\left.\left.+\Omega^2(2(1-3\omega_*)+6(1+\omega_*)(1-3\omega_*))\right\}\rho\rho_*\right]\\
%%%%%%%%%%%%%%%%%%%
\dot H_2 (y=1) =  &&\frac{-8\pi^2 G^2 (\Omega^{-2\eps}-\Omega^4)}{k^2 \kappa^2 v_0^2 \eps^2 (2+\eps)(1-\Omega^2)\Omega^2}\left[9\Omega^6(1-3\omega)(1+\omega)^2\rho^2 \right.\nonumber\\
&&+(1-3\omega_*)(1+\omega_*)(4(1-\Omega^2)+9(1+\omega_*))\rho_*^2\nonumber\\
&&\left.+\Omega^2\left\{(1-3\omega)(2(1+\omega_*)+6(1+\omega)^2+2(1+\omega)+3(1+\omega)(1+\omega_*)\right.\right.\nonumber\\
&&\left.\left.+\Omega^2(-2(1+\omega)(1-3\omega_*)+2(1+\omega_*)(1-3\omega_*)+3(1+\omega_*)(1-3\omega_*)(1+\omega)\right.\right.\nonumber\\
&&\left.\left.-4(1-3\omega)(1+\omega_*)+6(1+\omega_*)^2(1-3\omega_*))\right\}\rho\rho_*\right]\\
%%%%%%%%%%%%%%%%%%%%%%
\dot H_2 (y=0) =  &&\frac{-8\pi^2 G^2 (\Omega^{-2\eps}-\Omega^4)}{k^2 \kappa^2 v_0^2 \eps^2 (2+\eps)(1-\Omega^2)}\left[9(1-3\omega_*)(1+\omega_*)^2\rho_*^2 \right.\nonumber\\
&&-\Omega^4(1-3\omega)(1+\omega)(4(1-\Omega^2)+9\Omega^2(1+\omega))\rho^2\nonumber\\
&&\left.+\Omega^2\left\{(1-3\omega_*)\Omega^2(2(1+\omega_*)+6(1+\omega_*)^2+2(1+\omega)+3(1+\omega)(1+\omega_*)\right.\right.\nonumber\\
&&\left.\left.-2(1+\omega_*)(1-3\omega)+2(1+\omega)(1-3\omega)+3(1+\omega)(1-3\omega)(1+\omega_*)\right.\right.\nonumber\\
&&\left.\left.-4(1-3\omega_*)(1+\omega)+6(1+\omega)^2(1-3\omega)\right\}\rho\rho_*\right]
\eeqa

\section{Absence of a scalar field and infinite extra dimension}
In order to trust the perturbative approach which was used to derive the results presented in 
this paper,
we must convince ourselves that the same approach allows us to reproduce the well known 
results of RS-I and RS-II cosmology in the absence of a scalar field.  In this section, we 
will show that this is indeed the case.

In the absence of a stabilization mechanism, the configuration with two parallel branes 
lying at fixed positions along the extra dimension is generally unstable.  In order to 
look for static solutions, we must {\it impose} $b=$ constant from the start, which means
that any derivatives of $b$ are absent from the Einstein equations (\ref{EE}).  

Working to zeroth order in $\rho$, it is straightforward to show that:
\beqa
\label{zeronoscalar}
V_0 &=& -V_1 = {6 k \over \kappa^2}\nonumber\\
A_0 &=& k b_0 y,
\eeqa
so that we recover the expected fine-tuning between the branes' tensions.

The  (00) and (00-ii) combinations of the Einstein equations,
 linearized in $\rho$, can be solved in the same manner as in \cite{CF}.  
However, the absence of a scalar field means that the (55) equation no longer allows one
to solve for $\delta b_1 \over b_0$.  Rather, plugging the solutions for $\Psi_1$,$\Upsilon_1$,
$\left(\dot a_0 \over a_0\right)^2_1$ and $\left(\ddot a_0 \over a_0\right)_1$ into the (55) equation 
leads to the following constraint:
\beq
\Omega^4(\rho-3p)+\Omega^2(\rho_*-3p_*) = 0.
\eeq
The only way to satisfy this constraint without running into inconsistencies at higher order in 
$\rho$ is to set
\beq
\label{constraint1}
\rho = -\Omega^{-2} \rho_*; \quad
\omega_* = \omega = -1,
\eeq
where the second equation is found by demanding that ${d\over dt}\left(\rho \over \rho_*\right)$ be 
constant in time, as dictated by the first equation.  

In short, setting $\Phi$ and $\delta b$ to zero has made our system of
equations over-determined: the (00), (00-ii) and (55) Einstein
equations constitute a system of three equation for two unknowns.  The
way out of this  problem consists of turning one of our parameters
($\rho$ and $\rho_*$) into a variable to be expanded in powers of the
other.  We will therefore choose to write $\rho = \rho_1  + \rho_2 +
\dots$,  where the subscripts indicate the order in powers of $\rho_*$,
and $\rho_1$ is given by  (\ref{constraint1}).

We can then repeat the same steps (solving the first two Einstein
equations and plugging the  results in the (55) equation) for the
equations at second order in $\rho_*$ to find:
\beqa
\label{constraint2}
\rho_2 = \frac{-\kappa^2 (1-\Omega^2)}{12 k \Omega^4} \rho_*^2.
\eeqa
These results for the relation between $\rho$ and $\rho_*$ agree with
the exact results  presented in \cite{Kaloper}, when the latter are
expanded as a powers series in $\rho_*$.  

Imposing these constraints we find the Hubble rate on the Planck brane
to be
\beqa
H^2|_{y=0} = \frac{8\pi G}{3}(1-\Omega^2)\rho_* \left[1+\frac{\kappa^2}{12 k}\rho_*\right] + O(\rho_*^3)
\eeqa
with $8\pi G$ given by (\ref{8piG}).   Using (\ref{zeronoscalar}) and
the fact that $\Omega \rightarrow 0$ as $b_0 \rightarrow \infty$, we
see that we do indeed recover the expected RS-II behaviour (\ref{mfe})
when the second brane is taken to infinity.  We emphasize however that
this is only true insofar as $\rho_1$ and $\rho_2$ are written as in
(\ref{constraint1}) and (\ref{constraint2}).  If we had instead decided
to take the TeV brane out of the picture by setting $\rho = 0$ at the
start, we would not have gotten this result.  Specifically, there
would be a missing factor of $1/2$ in the second order correction, and
more importantly, the constraint equations
(\ref{constraint1}, \ref{constraint2}) would demand that $\rho_*=0$.

Suppose now that we {\it start out} with a single brane with an extra
dimension of infinite size.  The difference between this approach and
the one described above lies in the fact that setting $\rho=0$ also
sets the boundary conditions for $\Psi$ and $\Upsilon$ at $y=1$ to be
$\Psi(y=1) = \Upsilon(y=1) = 0$.  Here however, since the second brane
is explicitly absent from the setup, we only have one set of boundary
conditions on $\Psi$ and $\Upsilon$, i.e. these two variables are only
fixed on the single brane, located at $y=0$.  

We start by redefining the extra dimension's coordinate $y$ as 
\beq
\hat y = b_0 y,
\eeq
so that $\hat y$ goes from $0$ to $b_0$ as $y$ goes from $0$ to $1$
(and $b_0$ can be chosen to be infinite).  We also choose a gauge
where  the fluctuations in $b$ vanish.  To order $\rho^0$, the
equations tell us that
\beqa
A_0 = k \hat y\nonumber\\
V_0 = \frac{6k}{\kappa^2}
\eeqa
The order $\rho^1$ Einstein equations will again look like those of
\cite{CF}.  Even though we only have one set of boundary conditions for
$\Psi_1$ and $\Upsilon_1$, we can still solve the first two Einstein
equations, leaving $\Psi_1(\hat y = b_0)$ and $\Upsilon_1(\hat y =
b_0)$ unspecified.  If we then plug the results in the (55) equation,
we find the following constraint:
\beq
\left(4\Psi_1 + \Upsilon_1\right)|_{\hat y = b_0} = \Omega^{-2}\left(4\Psi_1 +
\Upsilon_1\right)|_{\hat y = 0}
\eeq 
Taking this into account, we can write the Friedmann equations as:
\beqa
H^2_1 &=& \frac{2k}{(1-\Omega^2)}\left(\frac{\kappa^2}{6}
\rho_*-\Omega^4\Psi_1|_{\hat y = b_0}\right)\\
\dot H_1 &=& \frac{k}{6(1-\Omega^2)}\left(\kappa^2(-\Omega^2+3\Omega^2 
\omega_* -3-3\omega_* )\rho_* + 24\Omega^4 \Psi_1|_{\hat y = b_0}\right)
\eeqa
It would appear that we have complete freedom in choosing a value for
$\Psi_1(\hat y = b_0)$, but there are a couple of points to consider.  

Firstly, we expect our results to reproduce the standard Friedmann 
equations at $O(\rho_*^1)$.  One can easily see that with the following choice:
\beqa
\Psi_1(\hat y = b_0) = \frac{\kappa^2}{6}\rho_*
\eeqa
this requirement is satisfied.  Similarly, recovering the correct 
$O(\rho_*^2)$ term would require setting $\Psi_2$ to the appropriate 
value at $\hat y = b_0$.

Secondly, we know from \cite{Kaloper} that if the tension on the brane is larger than the static solution's value, there will be a horizon in the bulk at the (finite) value of $\hat y$ for which $a(\hat y) = 0$.  Given the form we have chosen for $a$, it is not immediately obvious how we can reproduce this behaviour.  However, if we expand $e^{-\delta A_1}$ as $1 - \delta A_1$, then in the limit where we neglect any higher order in $\rho$ contributions, $a$ will vanish when $\delta A_1 = 1$.  (Notice that this is also the point at which our perturbative approach breaks down completely).  
Choosing $\Psi_1(\hat y = b_0)$ in the manner mentioned above, and recalling that 
\beqa
\delta A_1 (\hat y) = \int^{\hat y}_0 \Psi_1(\hat y)\, d\hat y
\eeqa
we find that $\delta A_1 (\hat y) = 1$ at the position
\beq
\hat y_{a=0} = \frac{1}{2k}\ln\left(1+\frac{12k}{\kappa^2 \rho_*}\right)
\eeq
which corresponds precisely to what we would find using the exact solutions of \cite{Kaloper}. 

One more subtlety remains.  If we solve for $\delta N_1$, we can see 
that there will be a point in the bulk where $n$ vanishes.  
(In the same manner as $a$ vanishes, {\it i.e.,} where $\delta N_1 (\hat y) 
= 1$).  In general, this point and the one where $a$ vanishes do not 
coincide.  Indeed, it can be shown that $\delta N_1 = 1$ at the position
\beq
\hat y_{n=0} = \frac{1}{2k}\ln\left(1+\frac{12k}{\kappa^2 \rho_*}+3(1+\omega_*)(e^{2k}-1)\right).
\eeq
Intuitively, one might expect that $a=0$ corresponds to a geometrical 
singularity while $n=0$ corresponds to a horizon.   We would then
demand that $\hat y_{n=0} \leq \hat y_{a=0}$ in order to avoid the
appearance of a naked singularity in the bulk.  One can readily see
that this would impose $\omega_* \leq -1$.  However, as we stated in
the main text, examining  the behaviour of the Riemann tensor at these
special values of $\hat y$ shows no evidence of curvature singularities
in the bulk.

To summarize, we have shown in this section that our perturbative
approach allows us to:

a) reproduce the expected RS-II result as a limit of unstabilized RS-I,
as long as $\rho$ is treated as a dependent rather than a free parameter;

b) reproduce the expected RS-II results starting from a single brane
setup, as long as we fix the values of $\Psi$ and $\Upsilon$ at $\hat y
= b_0$ appropriately;

c) calculate the points in the bulk at which $a$ and $n$ vanish and
find a result which matches the exact one.

\end{document}